\documentclass[a4paper,UKenglish]{oasics}

\usepackage{pifont} 
\usepackage{float}

\lstdefinelanguage{codeTTN}
{
        basicstyle=\ttfamily\footnotesize,
        sensitive=true,
        showstringspaces=false,
        numberblanklines=true,
        showspaces=false,
        breaklines=true,
        showtabs=false,
		numbers=left,
		numberstyle=\footnotesize,
		xleftmargin=15pt
}
\lstnewenvironment{code}{\lstset{language=codeTTN}}{}

\newcommand{\checkK}{\color{green}\checkmark}
\newcommand{\cross}{\color{red}\hspace{-3pt}\ding{55}}
\newcommand{\bigexclaim}{\color{yellow}$\bigtriangleup$\hspace{-5.6pt}!}

\title{Automatic Test Generation for Space}

\author[1]{Ulisses Ara\' ujo Costa}
\author[2]{Daniela da Cruz}
\author[3]{Pedro Rangel Henriques}
\affil[1]{VisionSpace Technologies\\
  Rua Alfredo Cunha, 37, Matosinhos, Portugal\\
  \texttt{ucosta@visionspace.com}}
\affil[2]{Department of Informatics, University of Minho\\
  Campus de Gualtar, 4710-057, Braga, Portugal\\
  \texttt{danieladacruz@gmail.com}}
\affil[3]{Department of Informatics, University of Minho\\
  Campus de Gualtar, 4710-057, Braga, Portugal\\
  \texttt{pedrorangelhenriques@gmail.com}}
\authorrunning{U. A. Costa et. al}

\Copyright[nc-nd]
          {Ulisses Ara\' ujo Costa and Daniela da Cruz and Pedro Rangel Henriques}


\serieslogo{}
\volumeinfo
  {Alberto Sim{\~o}es and Ricardo Queir{\'o}s and Daniela da Cruz}
  {3}
  {1st Symposium on Languages, Applications and Technologies}
  {21}
  {1}
  {185}
\EventShortName{SLATE'12}
\DOI{10.4230/OASIcs.SLATE.2012.185}

\begin{document}
\maketitle
\begin{abstract}
The European Space Agency (ESA) uses an engine to perform tests in the
Ground Segment infrastructure, specially the Operational Simulator.
This engine uses many different tools to ensure the development of
regression testing infrastructure and these tests perform black-box
testing to the C++ simulator implementation.
VST (VisionSpace Technologies) is one of the companies that provides
these services to ESA and they need a tool to  infer automatically tests from the existing C++ code, instead of writing manually scripts to perform tests.
With this motivation in mind, this paper explores automatic testing approaches and tools in order to propose a system that satisfies VST needs.
\end{abstract}

\section{Introduction}\label{introd}
Since ever, every industry use testing methods to discover problems in early stages of the development process to improve
the products quality, and software industry is not an exception. Miller\cite{miller} describe the utility
of software testing as:

\begin{quotation}
The general aim of testing is to affirm the quality of software systems by systematically
exercising the software in carefully controlled circumstances.
\end{quotation}

In the most recent period of software history the integration of
software testing as an important step in the process of
software development opened up to the origin of \textit{xUnit}\cite{xunit}
tools and Agile software development.
Also, ESA started to use manual written tests as a part of their
software development processes.\\
Using  manual written tests is tedious, time consuming and error-prone.
Lots of functions/methods need full code coverage and this practice
leads to incomplete test suites;
as it is hard to create tests that cover specific code paths, many
hidden bugs can be left.
Many times a supervision leaded by the developer
is needed to assure that the right paths in the code are being tested,
specially regarding black-box testing.\\
Nowadays we start to observe a rapid increase in the automatic test
generation field.

\subsection{Goals}
This document correspond to the first milestone in the author's dissertation (developed under a partnership agreement between UM and VST) aimed at producing a tool
that is able to automatically generate interesting testcases for the C++ ESA's Operational Simulator.\\
This document reviews the most studied techniques
and the tools that implement them in order to choose the best set of
suitable techniques to incorporate in an automatic
testing generator to the Ground Segment infrastructure, specially the
Operational Simulator at ESA.\\
Two different techniques emerge for different purposes, Structural
Techniques and Functional Techniques,
known respectively as White-box\cite{stt} testing and Black-box\cite{black} testing.
Functional testing is the most common at ESA, because of the
calculation complexity behind the Operational Simulators.\\
A brief discussion will be presented regarding White-box testing vs. Black-box
testing and then some automatic generation techniques will be discussed in more detail.
Furthermore the potential of the described tools will be explained, and how they can help
on solving the problem VST has nowadays. First of all an explanation about the Operational Simulator Infrastructure will be provided.

\subsection{Operational Simulator Infrastructure}
ESA's Operational Simulator called Simulation Infrastructure for the Modeling of SATellites (SIMSAT) is a satellite simulator that model and simulate
the behavior of satellites in order to allow operators\footnote{Operators are responsible for the operation of the satellite after its launch.} train more effectively 
and help them to define the satellites' operational processes.

The simulator consists of operational models of the various internal components of the satellite from their main computer to its payload (instruments aboard the satellite),
which interact with each other and thus define the behavior of the satellite.
VST has participated in the development of tests to validate the operational simulator.
The development of these simulators is based on operating rules simulation of
ESA -- Simulation Model Portability (SMP)\footnote{SMP is based on the ideas of component-based design and Model Driven Architecture (MDA)
as promoted by the Object Management Group (OMG) and is based on the open standards of UML and XML.
One of the basic principles is the separation of the platform specific and platform independent aspects of the simulation model.
This protects the investments in the model from changes in technology by defining the model in a platform independent way, which can then be mapped into different technologies.
Further the SMP specification provides standardised interfaces between the simulation models and the simulation run-time environment for common simulation services as well as a
number of mechanisms to support inter-model communication.\cite{1A,2A,3A,4A,5A}}, as well as in infrastructure SIMSAT simulation.
This standard is infrastructure agnostic of any space specific model, so any other needs of simulation can be used, such as defense, transport, energy, etc.\\

Here is a brief description of each component in SIMSAT\footnote{More information in: http://www.egos.esa.int/portal/egos-web/products/Simulators/simsat/intro-sim.html}:
\begin{description}
\item[SIMSAT Kernel] this is a generic simulation infrastructure providing the framework for the running of space systems simulators.
\item[SIMSAT Man-Machine Interface (MMI)] this is a generic Graphical User Interface enabling the user interaction with the simulator's components.
\item[Ground Models] this is a family of SIMSAT compatible models enabling a realistic simulations of all ground systems between the spacecraft (or spacecraft model) and the control centre at European Space Operations Centre (ESOC).
\item[Emulator Suite] On-board Processor Emulators support the execution in satellite simulators of the real flight software.
\item[Generic Models] a set of generic space models that ease the developments of the spacecraft models used in operational simulators.
\item[Ground Systems Test and Validation Applications (GSTV)] this is a family of test simulators that are based on the generic simulators infrastructure components listed above and are able to support the different levels of testing of ground infrastructure systems.
\end{description}

Moreover the SIMSAT Kernel is made up of several components\footnote{More information in: http://www.egos.esa.int/portal/egos-web/products/Simulators/SIMSAT/}:
\begin{description}
\item[Scheduler] is responsible for the co-ordination and processing of all events within the Simulation Kernel. An event on the schedule identifies an action that needs to be performed at a specified point in simulated time.
\item[Mode Manager] is the simulation state machine. The Simulation has a number of operational modes, which control the operation of the simulation.
\item[Time-Manager] is responsible for maintaining and providing models and the MMI with the correct simulation-Time. It provides time in four formats, Simulation-Time, Epoch-Time, Zulu-Time and Correlated Zulu-Time. this is a family of SIMSAT compatible models enabling a realistic simulations
\item[Logger] supports the recording of Kernel or model events that occur during a simulation. The log in which the current simulation messages are written is called the active log. The logger also provides a view of the simulation event history in an MMI during a simulation session.
\item[Visualization manager] is responsible for making the values of both model and Kernel data items available for display in an MMI.
\item[State-vector manager] is responsible for the saving and restoring of the state of the simulation. Its main purpose is to allow the Simulation State, at any point in the simulation, to be saved. This allows the user to return to an earlier simulation scenario.
\item[Command handler] is responsible for the reception and execution of Kernel and user defined commands.a set of generic space models that ease the developments of the spacecraft models used in operational.
\item[Command procedure] interpreter is responsible for the interpretation of command procedures. A command procedure contains Kernel and User defined simulator commands and supports a procedural language to control the flow of these commands. The execution of command procedures is controlled directly from the MMI.
\end{description}

Right now, to be able to perform tests in the Operational Simulator, in order to validate SIMSAT, VST Engineers need to write scripts that
perform simulations and validate the results using GUI interfaces (SIMSAT MMI). This job can be tedious and difficult to replicate.\\
So a first solution will have to go through a preliminary study of the tools
that currently exist with which we can generate tests automatically.
By studding these tools we do not hope to find the perfect solution, but combine techniques to obtain an optimal solution to improve VST work.

\subsection{White-box vs Black-box testing}
In this subsection is discussed the two most common approaches for testing: White-box and Black-box testing.\\
In White-box testing the tester needs to understand the internals of
the code to be able to write tests for it.
The goal of selecting test cases that test specific parts of the code
is to cause the execution of specific spots in the software, such as
statements, branches or
paths.
This technique consists in analyzing statically a program, by reading
the program code and using symbolic execution techniques to simulate
abstract program
executions in order to attempt to compute inputs to drive the program
along specific execution paths or branches, without ever executing the
program. Control Flow based testing approach can be useful to analyze all the
possible paths in the code and write unit tests to cover multiple paths.
The CFG (Control Flow Graph) of the program can be built,
test inputs can be generated to make any path execute regarding a given criterion:
Select all paths;
Select paths to achieve complete statement
coverage\cite{stt,Ntafos:1988:CST:630792.631017};
Select paths to achieve complete branch coverage\cite{Roper1994,stt};
or Select paths to achieve predicate
coverage\cite{stt,Ntafos:1988:CST:630792.631017}.

Data Flow Testing is designed into looking at the life cycle
(creation, usage and destruction) of a particular
piece of data and observe how it is used along the CFG, this ensures
that the number of paths is always finite\cite{dataflow}.\\

Opposite to White-box testing, Black-box testing is based on
functionality, so the tester observes a system based
on its functional contracts and writes the pairs of inputs and the
expected outputs.
This approach is used for unit testing of single methods/functions,
integration testing
of combinations of the methods/functions, or even final system testing.\\

This document is organized as follows.
In Section~\ref{testingapproaches} the important testing approaches in
use---Specification-based testing and Constraint-based generation---are briefly
revisited and, for each one,  the most relevant tools are identified.
In Section~\ref{testingtools} some of the tools referred are
experimented in order to be compared.
Our proposal for a test generation system is introduced in
Section~\ref{proposal}.
The document is concluded in Section~\ref{Concl}.

\section{Testing Tools Approaches}\label{testingapproaches}
In this section, a study of the most recent tools that use Specification-based, Constraint-based, Grammar-based and Random-based tests generation
approaches for the most popular languages - C, JAVA and C\# will be presented.

\subsection{Specification-based Generation Testing}
Specification Based Testing refers to the process of testing a program based on what its specification or model says its behavior should be.
In particular, can be generated test cases based on the specification of the program's behavior, without seeing an implementation of the program. So this clearly a
way of Black-box testing.\\
With this technique the testing phase and development phase can be started in parallel, we do not need the implementation
to start the development of test cases. The only thing needed is the functional contracts and/or oracles\footnote{A test oracle determines whether or not the results of a test execution are correct\cite{Peters95generatinga}.} for each function/method.\\
Since the 90's there have been some effort into using specifications to try to generate test cases such as Z specifications
\cite{Horcher95improvingsoftware,Stocks:1996:FST:239916.239918}, UML statecharts\cite{Offutt:1999:GTU:1767297.1767341},VDM\cite{Aichernig99automatedblack-box}
or ADL specifications\cite{Sankar94specifyingand}.
These specifications typically do not consider structurally complex inputs and these tools do not generate JUnit test cases.
Nowadays there are some tools out there that can perform Specification-based Testing approach:

\begin{description}
\item[Conformiq] is a commercial Tool Suite that generates
human-readable test plans and executable test scripts from Java code, state charts and UML\footnote{See more at: \url{http://www.conformiq.com/products.php}}.
\item[MaTeLo] stands for Markov Test Logic and is a commercial tool
that generates test sequences from a collection of states, transitions, classes of equivalence, types, sequences, global variables and test oracles
using their user interface\footnote{See more at: \url{http://www.all4tec.net/index.php/All4tec/matelo-product.html}}.
\item[Smartesting CertifyIt] is a commercial tool that generates test cases from a functional model, as UML\footnote{See more at: \url{http://www.smartesting.com/index.php/cms/en/product/certify-it}}.
\item[T-Vec] is a commercial tool that generates test cases from modeling tools available from T-VEC or third-party vendors\footnote{See more at: \url{http://www.t-vec.com/}}.
\item[Rational Tau] is an IBM commercial tool that provides automated error checking, rules-based model checking, and a model-based explorer using
UML\footnote{See more at: \url{http://www-01.ibm.com/software/awdtools/tau/}}.
\end{description}
The relevant ones or the recent open-source ones will be discussed.

\subsubsection{Spec Explorer}
This is a Microsoft model-based testing that uses one software modeling languages, the AsmL (Abstract State Machine Language).
This modeling language provides the foundations of the Spec Explorer\footnote{See more at: \url{http://research.microsoft.com/en-us/projects/specexplorer/}} tool
and Spec\# that is a formal language for API contracts (influenced by JML, AsmL, and Eiffel), which extends C\# with constructs for non-null types,
pre-conditions, post-conditions, and object invariants\footnote{See more at: \url{http://research.microsoft.com/en-us/projects/specsharp/}}.
These tool is already available to users and is in a very mature phase.\\
\indent The user of Spec Explorer writes a model of the system and sets the possible values for some properties in his code, furthermore the user also provides a scenario.
These scenarios are simple sets of calls to methods without their parameters (remember that this is Spec Explorer job).
Then Spec Explorer will generate a visual graph where each node represents a state of the system and the arrows represent a call to some method.
It searches throw all possible sequences of methods invocation that do not violate the contracts (pre, pos conditions) and
that are relevant to a user-specified set of test properties. After that we can generate from this visual graphs the unit tests (the arrows) and the
test cases (a graph).

\subsubsection{JMLUnit}
JMLUnit\cite{Cheon04thejml} is a tool that automates the generation of oracles for JAVA testing classes. This tool
monitors the specified behavior of the method being tested to decide whether the test passed or failed.
This monitoring is done using the formal specification language runtime assertion checker.
The main idea behind these tools is to translate the pre- and post-conditions methods into the code of the testing method.\\
The pre-conditions became the criteria for selecting test inputs, and the post-conditions provided the properties to check for
test results. So, the post-conditions became the test oracles.\\
This tool uses the JML\cite{Burdy03anoverview} specification language to annotate JAVA methods code with pre- and post-conditions and
automatically generate JUnit test classes from JML specifications.

\subsubsection{TestEra}
TestEra\cite{testera} can be used to perform automated specification-based testing of
JAVA programs. This framework requires as input a JAVA method, a formal specification\footnote{Specifications are first-order logic formulae.}
of the pre and post-conditions of that method, and a bound that limits the size of the test cases to be generated.\\
With the pre-condition it automatically generates all non-isomorphic test inputs up to the given bound.
It executes the method on each test input, and uses the method post-condition as an oracle to check the correctness of each output. This tool
uses Alloy's\footnote{Alloy is a first-order declarative language based on sets and relations. The Alloy Analyzer is a fully
automatic tool that finds instances of Alloy specifications: an instance
assigns values to the sets and relations in the specification such that
all formulae in the specification evaluate to true.} SAT system to analyze first-order  formulae.
The authors claim that have used TestEra to check several JAVA programs including an architecture for
dynamic networks, the Alloy-alpha analyzer, a fault-tree analyzer, and methods from the JAVA Collection Framework.

\subsubsection{Korat}
Korat\cite{Boyapati02korat:automated} is a mature framework for automated testing structurally complex inputs of JAVA programs.
Given a formal specification for a method, Korat\footnote{See more at: \url{http://korat.sourceforge.net/}} uses the method pre-condition
to automatically generate all (non-isomorphic) test cases up to a given small size.
Korat then executes the method on each test case, and uses the method post-condition as a test oracle to check the correctness of each output.\\
To be able to generate test cases for a method, Korat uses a predicate and a bound on the size of its inputs,
Korat generates all (non-isomorphic) inputs for which the predicate returns $true$.
Korat generates all the possible input spaces regarding the predicate and monitor the predicate's executions to be able to prune large portions of the search space.\\
\indent The writing of a predicate is done using JAVA language and in most cases can be written the first thing that cames to programmer's head to restrict the input space.
But for more complex structures it is better to understand how the matching algorithm work to be able to write a fast verifiable predicate.\\
Unfortunately the test derivation tool using Korat (that also uses JML) is not available to the public.

\subsection{Constraint-based Generation Testing}
Constraint Based Testing\cite{DeMillo91constraint-basedautomatic} can be used to select test cases satisfying specific constraints by
solving a set of constraints over a set of variables. The system is described using constraints and these can be solved by SAT solvers.\\
Constraint programming can be combined with symbolic execution, regarding this approach a program is executed symbolically,
collecting data constraints over different paths in the CFG, and then solving the constraints and producing test cases from there.
There are some tools out there, like:

\begin{description}
\item[Euclide] for verifying safety properties over C code using ACSL annotations, CPBPV for program verification.
\item[OSMOSE] a tool that uses concolic execution and path-based techniques over machine code.
\item[GATeL] for Lustre language to generate test sequences\footnote{See more at: \url{http://www-list.cea.fr/labos/gb/LSL/test/gatel/index.html}}.
\end{description}

Here two tools will be explained, one proprietary and other academic.

\subsubsection{Pex} Pex\cite{Tillmann:2008:PWB:1792786.1792798} is an automatic white-box test generation tool for .NET. Starting from a
method that takes parameters, Pex performs path-bounded model-checking
by repeatedly executing the program and solving constraint systems to obtain inputs that will steer the program along different execution paths.
This uses the idea of dynamic symbolic execution\cite{Tillmann06unittests}. Pex uses the theorem prover and
constraint solver Z3\footnote{See more at: \url{http://research.microsoft.com/en-us/um/redmond/projects/z3/}} to reason about the feasibility of execution paths, and
to obtain ground models for constraint systems.\\
Pex came with Moles that helps to generate unit tests. These tools together are able to understand the input (by analyzing branches in the code:
declarations, all exceptions throws operations, if statements, asserts and .net Contracts). With this information Pex uses Z3 constraint solver to
produce new test inputs which exercise diferent program behavior.\\
The result is an automatically generated small test suite which often achieves high code coverage.\\
Pex can be used in a project, class or method (which makes it a very helpful and versatile tool). After the analysis process the "Pex Explorarion Results" shows
the $input \times output$ pairs selected for each test case for the method, here it also shows the percentage of the test coverage.

\subsubsection{PathCrawler} This is an academic tool based on dynamic and static analysis\cite{Williams05pathcrawler:automatic}, 
it uses constraint logic programming to generate the Test-cases. PathCrawler\footnote{See more at: \url{http://www-list.cea.fr/labos/gb/LSL/test/pathcrawler/index.html}} executes an instrumented function for each function under test
with the generated inputs, it preserves this information to not cover the same path.\\
This tool supports assertions in any point in the code and pre-conditions regarding the input values.

\subsection{Grammar-based Generation Testing}
In this approach inputs to a system under test are defined by a context-free grammar. The language of the grammar contains all possible test cases.
Using this approach to describe the syntax of the input to the system under test proves to be very helpful to test
network protocols\cite{tal:syntax-based,kaksonen2001functional} and parsers and compilers\cite{1994-burgess,Burgess_Saidi_1996}.

\subsubsection{ASTGen}
ASTGen\cite{Daniel:2007:ATR:1287624.1287651} is a JAVA framework that automates testing of refactoring engines: generation of test inputs
and checking of test outputs. The main technique is an iterative generation of structurally complex test inputs.
ASTGen\footnote{See more at: \url{http://mir.cs.illinois.edu/astgen/}} allows developers to write imperative generators whose executions
produce input programs for refactoring engines. More precisely, ASTGen
offers a library of generic, reusable, and composable generators that produce abstract syntax trees (ASTs).\\
So, ASTGen ensures the production of test inputs instead of the developer produce them. The developer needs to write a generator whose execution
produces thousands of programs with structural properties that are relevant for the specific refactoring being tested. This tool has found
21 bugs in Eclipse and 26 bugs in Netbeans applications.

\subsection{Random-based Generation Testing}
In the random testing approach, test inputs are selected randomly from the input domain of the system.
To have a random testing suite first we must identify the input domain, after that select test inputs independently from the domain,
then the system under test is executed on these inputs, the results are compared to the system specification, an oracle.\\
Random testing gives us an advantage of easily estimating software reliability from test outcomes.
Test inputs are randomly generated according to an operational profile, and failure times are recorded.
The data obtained from random testing can then be used to find bugs or non expected behaviors.\\
\indent The main problem regarding random generation is the problem of the coverage, it is possible that it will not be broad enough. And furthermore it can be
too sparse to actually test specifics parts of the program. Either way, this technique proves to be very effective for testing compilers.

\subsubsection{Csmith}
Csmith\cite{Yang:2011:FUB:1993316.1993532} is a black-box random tests generator that is able to generate C programs
conform to the C99\footnote{See more at: \url{http://www.open-std.org/jtc1/sc22/wg14/www/docs/n1256.pdf}} standard. This is a very recent tool that already discover
more than 195 bugs in LLVM and 79 bugs in GCC. With Csmith we are able to generate random programs with unambiguous meanings (undefined behavior or 
unspecified behavior). Does not attempt to generate terminating program, so they use timeouts for long time consuming generated programs.
And the main supported features right now are: Arithmetic, logical, and bit operations on integers, Loops, Conditionals, Function calls, Const and volatile,
Structs and Bitfields, Pointers and arrays, Goto, Break and continue. The generation of code regarding this features can be tuned using the command line program.

\subsubsection{QuickCheck for JAVA}
QuickCheck was originally a combinator library for the Haskell\footnote{See more at haskell.org} programming language\cite{Claessen:2000:QLT:357766.351266}.
Later on QuickCheck philosophy spread to other programming languages like: JAVA, Erlang, Perl, Ruby and JavaScript.\\
QuickCheck works by generating high amounts of data (within the method domain) and checking it against a given property,
it is expected to create a wide range of the input domain, thus increasing the chances of giving more test coverage.

\section{Using the tools}\label{testingtools}
After introducing the theory and the techniques that support each tool, some of the tools will be demonstrated in action, resorting to small but illustrative examples
on how each tool can help us to find good test cases.\\

\subsection{PathCrawler}
Concerning the first case  a simple example will be used based on a function that performs a multiplication, creating a simple branch on the code.
\begin{code}
typedef struct s {
    int x;
    int y;
}Point;

int Multiply(Point p) {
    if(p.x * p.y == 42) return 1;
    else return 0;
}
\end{code}
Pointers were tried instead of coping the structure as a parameter to $Multiply$ function, but PathCrawler was not able to run.

Nevertheless, PathCrawler was able to give a full coverage for this simple function as you can see in Table \ref{tab:mul}.

\begin{table}[!ht]
\renewcommand{\arraystretch}{1.3}
\setlength{\tabcolsep}{10pt}
\caption{Output Table for $Multiply$ function using PathCrawler}
\label{tab:mul}
\centering
\noindent \begin{tabular}{|c|c|c|}\hline
Result & p & return value\\\hline
\checkK & Point\{x=1,y=42\} & 1 \\\hline
\checkK & Point\{x=177407,y=109471\} & 0 \\\hline
\end{tabular}
\end{table}

Regarding our second example a function that performs a binary search in order to find if a number is in a given range (between two bounds).

\begin{code}
int BSearch(int x, int n) {
    return BinarySearch(x, 0, n); 
}
	
int BinarySearch(int x, int lo, int hi) {
    while (lo < hi) {
        int mid = (lo+hi)/2;
        pathcrawler_assert(mid >= lo && mid < hi);
        if (x < mid) { hi = mid; }
		else { lo = mid+1; }
    }
    return lo; 
}
\end{code}
A function that PathCrawler gives to us has been used: $pathcrawler\_assert$, this function can be used at any location in the
program under test, and will force PathCrawler to generate test cases to cover both the case where its argument is true and the case where it is false.
This feature may be seen as another way to write an oracle.\\
The results were interesting: 31 covered paths and 44 infeasible paths and the test was interrupted by PathCrawler,
because PathCrawler reach the maximal test session time (the user can increase this number, but for this example is left the default value).\\
A further analysis of the results demonstrated that 28 out of the 44 infeasible paths discovered appeared when PathCrawler tried to
do the assertion in line 8. No pre-condition was written, so PathCrawler does not know that this is a pre-condition
for $BinarySearch$ function:  $lo\leq~x<hi$. In Table \ref{tab:bsearch} is shown some of the test inputs generated for this example.

\begin{table}[!ht]
\renewcommand{\arraystretch}{1.3}
\caption{Output Table for $BSearch$ function using PathCrawler}
\label{tab:bsearch}
\centering
\noindent \begin{tabular}{|c|c|c|c|}\hline
Result & x & n & return value \\\hline
\checkK & -189424 & -140714 & 0 \\\hline
\checkK & 157819 & 0 & 0 \\\hline
\checkK & 1 & 1610612736 & 2 \\\hline
\checkK & 2 & 805306368 & 3 \\\hline
\checkK & 11 & 1610612736 & 12 \\\hline
\end{tabular}
\end{table}

PathCrawler was tried with the following function that calculates the year of the $n^{th}$ day after 1980-01-01.

\begin{code}
int IsLeapYear(int year) {
  return (year % 4 == 0) && ((year % 100 != 0) || (year % 400 == 0));
}
int FromDayToYear(int day) {
  int year = 1980;

  while (day > 365) {
    if (IsLeapYear(year)) {
      if (day > 366) {
        day -= 366;
        year += 1;
      }
    } else {
      day -= 365;
      year += 1;
    }
  }
  return year;
}
\end{code}

The result was unexpectedly $unknown$. PathCrawler was unable to trace even one path in our code, the number of $k$-path's could
be increased but with no success for this example.

\subsection{Pex}
Regarding Pex, we used the same examples shown previously adapted to C\# language.
Because C\# is a more expressive language than C our examples will be improved with some other OO and C\# specific features like Exceptions and Debug.Assert calls.
In fact Pex can also support a lot more features that are present in C\# language like .NET Contracts and many more.\\
This is the simple implementation of a 2D $Point$ class that has been created to have special behavior, under a certain condition
$x \times y \equiv 42$ it is supposed to throw an exception.

\begin{code}
public class Point {
  public readonly int X, Y;
  public Point(int x, int y) { X = x; Y = y; }
}

public class Multiply {
  public static void multiply(Point p) {
    if (p.X * p.Y == 42)
        throw new Exception("hidden bug!");
  }
}
\end{code}

So, as was described earlier, Pex will try to generate such input as it is possible (in a given amount of time) to traverse all the paths inside the code.
The output table can be seen in Table \ref{tab:point}, with the inputs and outputs that Pex found to ensure a full coverage of the code.

\begin{table}[!ht]
\renewcommand{\arraystretch}{1.3}
\setlength{\tabcolsep}{1pt}
\caption{Output Table for $multiply$ method using Pex}
\label{tab:point}
\centering
\noindent \begin{tabular}{|c|c|c|c|}\hline
Result & p & Output/Exception & Error Message\\\hline
 &  &  & Object ref. not set \\
\cross & null  & NullReferenceException & to an instance \\
 &  &  & of an object.\\\hline
\checkK & new Point\{X=0,Y=0\} & &\\\hline
\cross & new Point\{X=3,Y=14\} & Exception & hidden bug!\\\hline
\end{tabular}
\end{table}

Pex was successful to reach the $Exception$ path inside the code. Of course this is not always possible, since sometimes the functions inside
the $if$ statement does not have inverse function.\\

Pex can also be very helpful checking assertions and contracts in .net code. A binary search algorithm was written and an assertion was also written in
the middle of our code.

\begin{code}
public class Program {
  public static int BSearch(int x, int n) {
    return BinarySearch(x, 0, n);
  }
  static int BinarySearch(int x, int lo, int hi) {
    while (lo < hi) {
      int mid = (lo+hi)/2;
      Debug.Assert(mid >= lo && mid < hi);
      if (x < mid) { hi = mid; } else { lo = mid+1; }
    }
    return lo;
  }
}
\end{code}

Pex was able to generate an input that could not pass in the assertion inerted in our code, as can be seen in Table \ref{tab:binary}.

\begin{table}[!ht]
\renewcommand{\arraystretch}{1.3}
\setlength{\tabcolsep}{1pt}
\caption{Output Table for $BSearch$ method using Pex}
\label{tab:binary}
\centering
\noindent \begin{tabular}{|c|c|c|c|c|}\hline
Result & x & n & result & Output/Exception \\\hline
\checkK & 0 & 0 & 0      & \\\hline
\checkK & 0 & 1 & 1      & \\\hline
\checkK & 0 & 3 & 1      & \\\hline
\cross & 1073741888 & 1719676992 & & TraceAssertionException \\\hline
\checkK & 1 & 6 & 2      & \\\hline
\checkK & 50 & 96 & 51      &\\\hline
\end{tabular}
\end{table}

Now we have a more complex example, a function that returns the year of the $n^{th}$ day after 1980-01-01.
Pex was able to generate some important test cases, but it has reached the limit amount of time to calculate interesting paths in the code,
this boundary prevents Pex from getting stuck when the program goes into
an infinite loop.

\begin{code}
public class Program {
  private static bool IsLeapYear(int year) {
    return (year % 4 == 0) && ((year % 100 != 0) || (year % 400 == 0));
  }
  public static void FromDayToYear(int day, out int year) {
    year = 1980;
    while (day > 365) {
      if (IsLeapYear(year)) {
        if (day > 366) {
          day -= 366;
          year += 1;
        }
      } else {
        day -= 365;
        year += 1;
      }
    }
  }
}
\end{code}

Pex was unable to discover the year for day $366$ and $7671$ as we can see in Table \ref{tab:leap}.
This problem occurred because Pex by default has a maximum number of conditions, this avoids never ending functions and still has a result from Pex.
In this particular case we could increment the number of $MaxConditions$: $[PexMethod(MaxConditions=10000)]$.

\begin{table}[!ht]
\renewcommand{\arraystretch}{1.3}
\caption{Output Table for $FromDayToYear$ method using Pex}
\label{tab:leap}
\centering
\noindent \begin{tabular}{|c|c|c|c|c|}\hline
Result & day & out year & Output/Exception\\\hline
\checkK & 0 & 1980 & \\\hline
\checkK & 367 & 1981 & \\\hline
\bigexclaim & 366 & & path bounds exceeded\\\hline
\checkK & 1023 & 1982 &\\\hline
\checkK & 2561 & 1987 & \\\hline
\checkK & 7874 & 2001 & \\\hline
\bigexclaim &  7671 & & path bounds exceeded\\\hline
\end{tabular}
\end{table}

\subsection{Korat}
Like was explained before, Korat generates a graphical representation of the structure instances that validates the property $repOK$. This property was written using JAVA code.\\
In order to test the freelly available version of Korat, a Doubly Linked List structure was created in JAVA.

\begin{code}
public class LinkedList<T> {
  public static class LinkedListElement<T> {
    public T Data;
    public LinkedListElement<T> Prev;
    public LinkedListElement<T> Next;
  }
  private LinkedListElement<T> Head;
  private LinkedListElement<T> Tail;
  private int size; 
}
\end{code}

\def\t#1#2#3#4{\langle#1 \ #2 : #3 \ : #4 \ \rangle}
\def\d#1#2#3{\langle#1 \ #2 :: #3 \ \rangle}
\newcommand{\subseteqL}{\mathbin{\subseteq\mkern-4mu\subseteq}}
\newcommand{\inL}{\mathbin{\in\mkern-4mu\in}}

Now the $repOK$ predicate method must be defined.
This predicate method will check that the tree doesn't have any cycles and that the number of nodes traversed from root matches the value of the field size.
First was defined the properties about this data structure. The most relevant ones are property \ref{eq:linked} in Figure \ref{fig:formulae} that
ensures the structure and property \ref{eq:uniq} that ensures our doubly linked list does not have repeated elements.\\
Consider $e,e_1,e_2 \in LinkedListElement$ and $i$ the index function: $i : LinkedListElement \rightarrow int$, that receives an element of $LinkedList$ and
returns the position of that element in the structure. Consider also three new functions:
\begin{enumerate}
\item $Head(l)$ being $l$ of type $LinkedList$ and meaning in Java code $l.Head$.
\item $Tail(l)$ being $l$ of type $LinkedList$ and meaning in Java code $l.Tail$.
\item $size(l)$ being $l$ of type $LinkedList$ and meaning in Java code $l.size$.
\end{enumerate}

As a matter of avoiding verbosity two symbols were defined ($\inL$ and $\subseteqL$, these symbols are used to define the $LinkedList$ invariants in Figure \ref{fig:formulae}):
\begin{enumerate}
\item $a \inL l$ being $a$ of type $LinkedListElement$ and meaning that $a$ is an element of the $LinkedList$ $l$.
\item $\{a,\ldots,z\} \subseteqL l$ meaning $a \inL l \wedge \ldots \wedge z \inL l$.
\end{enumerate}

\begin{figure*}[!Hb]
\begin{eqnarray}
\t \forall {l} {l \in LinkedList} {Head(l) \equiv null \vee Tail(l) \equiv null \Leftrightarrow size(l) \equiv 0}\\
\t \forall {l} {l \in LinkedList} {Tail(l).Next \equiv null}\\
\t \forall {l} {l \in LinkedList} {Head(l).Prev \equiv null}\\
\t \forall {l} {l \in LinkedList} {size(l) \equiv 1 \Leftrightarrow Head(l) \equiv Tail(l)}\\
\t \forall {l} {l \in LinkedList} {\t \forall {e_1,e_2} {\{e_1,e_2\} \subseteqL l} {\t \exists {e} {e \inL l} {e_1.Next \equiv e \wedge e_2.Prev \equiv e}}\label{eq:linked}}\\
\t \forall {l} {l \in LinkedList} {\t \forall {e_1,e_2} {\{e_1,e_2\} \subseteqL l} {e_1 \equiv e_2 \Rightarrow i(e_1) \equiv i(e_2)}\label{eq:uniq}}
\end{eqnarray}
\caption{Invariants for class $LinkedList$}
\label{fig:formulae}
\end{figure*}

We took the properties described in Figure \ref{fig:formulae} and use them to restrict the generation of structures as we can see in the following Java implementation code.
Note that we using short-circuiting, so we return $false$ as soon as we can. This way Korat will be able to generate faster the instances matching our criteria.

\begin{code}
public boolean repOK() {
  if(Head == null || Tail == null)
    return size == 0;
  if(size == 1) return Head == Tail;
  if(Head.Prev != null) return false;
  if(Tail.Next != null) return false;
  LinkedListElement<T> last = Head;
  Set visited = new HashSet();
  LinkedList workList = new LinkedList();
  visited.add(Head);
  workList.add(Head);
  while (!workList.isEmpty()) {
    LinkedListElement<T> current = (LinkedListElement<T>) workList.removeFirst();
    if (current.Next != null) {
      if (!visited.add(current.Next))
	    return false;
      workList.add(current.Next);
      if(current.Next.Prev != current) return false;
      last = current.Next;
    }
  }
  if(last != Tail)
    return false;
  return (visited.size() == size);
}
\end{code}

The last step was defining the finitization method, this way we tell Korat how to bound the input space.

\begin{code}
public static IFinitization finLL(int nodesNum, int minSize, int maxSize) {
  IFinitization f = FinitizationFactory.create(LL.class);
  IObjSet nodes = f.createObjSet(LinkedListElement.class, nodesNum, true);
  f.set("Head", nodes);
  f.set("Tail", nodes);
  f.set("size", f.createIntSet(minSize, maxSize));
  f.set("LinkedListElement.Next", nodes);
  f.set("LinkedListElement.Prev", nodes);
  return f;
}
\end{code}

\begin{figure}[!ht]
\centerline{
\subfloat[Instance with $2$ elements for $LinkedList$]{
\includegraphics[width=.3\textwidth]{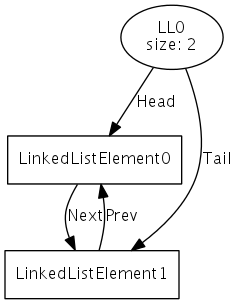}
\label{fig:inst1}
}
\hfil
\subfloat[Instance with $5$ elements for $LinkedList$]{
\includegraphics[width=.3\textwidth]{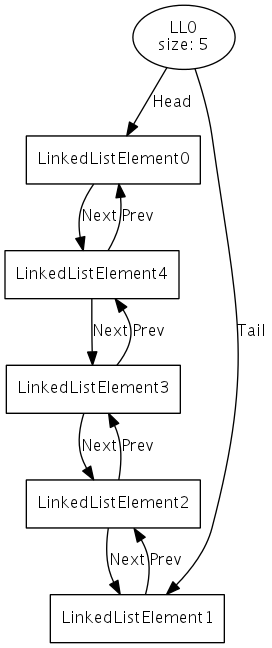}
\label{fig:inst2}}}
\caption{Examples of generated instances from Korat for $LinkedList$ class.}
\label{fig:insts}
\end{figure}

The properties in Figure \ref{fig:formulae} were taken and used to restrict the generation of structures using Java. So the $repOK$ method that receives
a $LinkedList$ structure and returns $Bool$ whenever this structure follows the invariants in \ref{fig:formulae} was defined.
Using this specification Korat generated the 2 structures shown in Figure \ref{fig:insts}. In Figure \ref{fig:inst1} with $2$ elements
and in Figure \ref{fig:inst2} an instance with $5$ elements.

\subsection{Summary}
\begin{figure}[!ht]
\centerline{
\subfloat[Example of Pex generated $LinkedList$ instance to test $Remove$ method]{
\includegraphics[width=.3\textwidth]{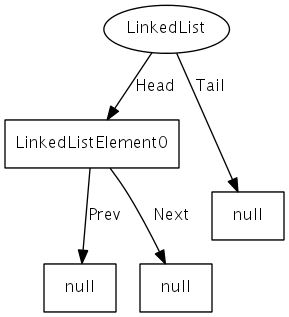}
\label{fig:pexinst1}
}
\hfil
\subfloat[Example of Pex generated $LinkedList$ instance to test $Find$ method]{
\includegraphics[width=.3\textwidth]{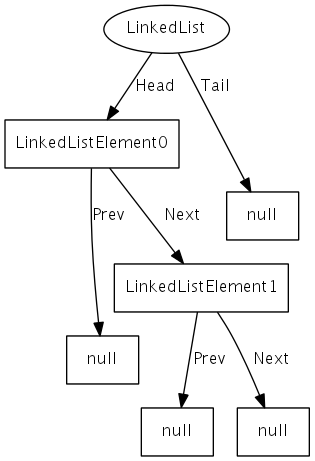}
\label{fig:pexinst2}}}
\caption{Examples of generated instances from Pex for $LinkedList$ class.}
\label{fig:pexG}
\end{figure}

After the experimental study of the selected tools, reported in the previous subsections, it was found that PathCrawler and Pex have different
approaches regarding testcase generation. PathCrawler seems to be a very efficient tool to discover multiple
infeasible paths in C code, because it uses a mix between static and dynamic analysis. When it finds a suitable input for a function it tries to execute
collecting all the executed paths in the code.
Pex on the other side just uses static execution and it is very efficient discovering all the feasible paths in C\# methods. Pex was also used
to perform testcase generation in C\# classes, but the generated instances are too simple to perform more interesting tests. The $LinkedList$ class was written
in C\# with many management methods implemented (Add, Remove, Find,\ldots). Pex generated very simple $LinkedList$'s structures to perform automatic test generation
for each implemented method. The problem is that the generated structures does not meet the properties about Doubly Linked Lists as it can be seen in Figure \ref{fig:pexG}.
Concerning Korat, this is The tool to generate complex data structures. The freely available part of Korat show potential in expressing rules to hedge
the automatic generation of data structures.\\
In Table \ref{tab:tabcmp} we can see a brief comparison between all the experimented and mentioned tools, a more detailed conclusion is addressed in Chapter \ref{Concl}.

\begin{table}[!ht]
\centering
\begin{tabular}{|m{2cm}|m{2cm}|m{2cm}|m{2cm}|m{2cm}|m{2cm}|}\hline
Name & Target Language & Black/White-box & Additional Input & Output & Comments\\\hline
\textbf{PathCrawler} & C & White-box (symbolic execution) & Test vectors & Constraints about the executed paths & Too Complex\\\hline
\textbf{Pex} & C\# & White-box (symbolic execution) & -- & Unit Tests & Poor generated data instances (objects)\\\hline
\textbf{Korat} & JAVA & Black-box & Invariants written in JAVA & Graphical form of data structures (using Alloy-GraphViz) & Powerful generating valid data instances\\\hline
\end{tabular}
\caption{Comparison of experimented and mentioned tools}
\label{tab:tabcmp}
\end{table}

\section{Generate Tests from Code+OCL}\label{proposal}
Since the Operational Simulator code is not familiar to us, regarding its implementation, it was decided to start solving this problem by inferring the UML+OCL from the existing code
to be able to work on a more abstract level rather than the implementation.
The idea is to extract tests from the inferred OCL, using the Partition Analysis described
in \cite{Benattou02generatingtest} and at the same time generate tests directly from the code, using symbolic execution to complement
the specification-based generation from OCL. The main goal is to extract as many tests as possible from a model and from the implementation 
to provide information to a feedback loop\cite{Xie03mutuallyenhancing}
test generation framework with two test prespectives, functional and structural, and from there be able to get a more refined set of tests.\\
A combination of both, symbolic execution from Pex and complex data generation from Korat, it will be designed and implemented to
generate more interesting inputs for the methods under testing.

\section{Conclusion}\label{Concl}
Looking for an efficient solution to automatically generate complete test sets for complex and critical C++ software,
the state-of-the-art approaches in the area were studied and along the document some tools were introduced from methodological and experimental perspectives.
Pex has proved to be a very powerful tool, aimed at offering a full coverage. However, the incapability for generating calling-methods sequences was a bit disappointing. 
With Microsoft's SpecExplorer we can already
manually call sequences of methods; maybe a combination of this feature with Pex would make Pex a perfect all-in-one testing tool regarding .NET automatic testing tools.
Concerning Korat, the expected improvement is just to write the invariants for a class instead of the $repOK$ method, or maybe infer these invariants 
from the existing code. Writing the $repOK$ method for very complex data structures requires some previous experience with Korat, but we think
this is not a weakness, since the tester quickly gets used to write the $repOK$ method in Korat. The only problem is that right now we can not fully automate the process
without human help.\\
\indent Considering the studied tools and thinking about a full automated test generation tool, a clever composition among between Pex to ensure the maximum possible coverage, 
Korat to generate all the valid data structures and an automatic tool to generate calls to methods combinations would be the perfect tool.\\

At the end, it was proposed  an approach based on the inference of tests from a Code+OCL.
\indent Concerning the OCL inference from C++ code, work will now be done on a tool that implements it.
For that purpose, Frama-C will be explored, as it is well known that this tool is able to infer pre- and post-conditions\cite{moy}
and interesting safety conditions from C source code.

\bibliography{slate12-ulisses}

\end{document}